\documentclass[twocolumn]{aastex631}
\usepackage{graphicx}
\usepackage{natbib}

\graphicspath{{./}{figures/}}
\received{2024 April 3}
\revised{2024 June 12}
\accepted{2024 July 1}
\submitjournal{ApJ Letters}

\newcommand\swinburne{Centre for Astrophysics and Supercomputing, Swinburne University of Technology, P.O. Box 218, Hawthorn, Victoria 3122, Australia}
\newcommand{\OzGravSwin}{OzGrav: The Australian Research Council Centre of Excellence for Gravitational Wave Discovery, Hawthorn VIC 3122, Australia}
\newcommand\CSIRO{Australia Telescope National Facility, CSIRO, Space and Astronomy, P.O. Box 76, Epping, NSW 1710, Australia}
\newcommand\MQ{Department of Physics and Astronomy and MQ Research Centre in Astronomy, Astrophysics and Astrophotonics, Macquarie University, NSW 2109, Australia}

\shorttitle{PPTA timing analysis of PSR~J0437$-$4715}
\shortauthors{Reardon, D. J. et al.}

\begin{document}

\title{The neutron star mass, distance, and inclination from precision timing of the brilliant millisecond pulsar J0437$-$4715}

\correspondingauthor{Daniel J. Reardon}
\email{dreardon@swin.edu.au}

\author[0000-0002-2035-4688]{Daniel J. Reardon}
\affil{\swinburne}
\affil{\OzGravSwin}

\author[0000-0003-3294-3081]{Matthew Bailes}
\affil{\swinburne}
\affil{\OzGravSwin}

\author[0000-0002-7285-6348]{Ryan M. Shannon}
\affil{\swinburne}
\affil{\OzGravSwin}

\author[0000-0003-1110-0712]{Chris Flynn}
\affil{\swinburne}
\affil{\OzGravSwin}

\author[0009-0002-9845-5443]{Jacob Askew}
\affiliation{\swinburne}
\affiliation{\OzGravSwin}

\author[0000-0002-8383-5059]{N.~D.~Ramesh Bhat}
\affiliation{International Centre for Radio Astronomy Research, Curtin University, Bentley, WA 6102, Australia}

\author[0000-0001-7016-9934]{Zu-Cheng Chen}
\affiliation{Department of Physics and Synergetic Innovation Center for Quantum Effects and Applications, Hunan Normal University, Changsha, Hunan 410081, China}
\affiliation{Institute of Interdisciplinary Studies, Hunan Normal University, Changsha, Hunan 410081, China}

\author[0000-0002-7031-4828]{Ma\l{}gorzata Cury\l{}o}
\affiliation{Astronomical Observatory, University of Warsaw, Aleje Ujazdowskie 4, 00-478 Warsaw, Poland}

\author[0000-0002-0475-7479]{Yi Feng}
\affil{Research Center for Astronomical Computing, Zhejiang Laboratory, Hangzhou 311100, China}

\author[0000-0003-1502-100X]{George B. Hobbs}
\affiliation{\CSIRO}

\author[0009-0001-5071-0962]{Agastya Kapur}
\affiliation{\CSIRO}

\author[0000-0002-0893-4073]{Matthew Kerr}
\affiliation{Space Science Division, US Naval Research Laboratory, 4555 Overlook Ave SW, Washington DC 20375, USA}

\author[0000-0002-2187-4087]{Xiaojin Liu}
\affiliation{Faculty of Arts and Sciences, Beijing Normal University, Zhuhai 519087, People`s Republic of China}

\author[0000-0001-9445-5732]{Richard N. Manchester}
\affiliation{\CSIRO}

\author[0000-0001-5131-522X]{Rami Mandow}
\affiliation{\MQ}
\affiliation{\CSIRO}

\author[0009-0001-5633-3512]{Saurav Mishra}
\affiliation{\swinburne}

\author[0000-0002-1942-7296]{Christopher J. Russell}
\affiliation{CSIRO Scientific Computing, Australian Technology Park, Locked Bag 9013, Alexandria, NSW 1435, Australia}

\author[0000-0003-3977-708X]{Mohsen Shamohammadi}
\affiliation{\swinburne}
\affiliation{\OzGravSwin}

\author[0000-0001-8539-4237]{Lei Zhang}
\affiliation{National Astronomical Observatories, Chinese Academy of Sciences, A20 Datun Road, Chaoyang District, Beijing 100101, People`s Republic of China}
\affiliation{\swinburne}

\author[0000-0002-9583-2947]{Andrew Zic}
\affiliation{\CSIRO}


\begin{abstract}
The observation of neutron stars enables the otherwise impossible study of fundamental physical processes. The timing of binary radio pulsars is particularly powerful, as it enables precise characterization of their (three-dimensional) positions and orbits.
PSR~J0437$-$4715 is an important millisecond pulsar for timing array experiments and is also a primary target for the Neutron Star Interior Composition Explorer (NICER). The main aim of the NICER mission is to constrain the neutron star equation of state by inferring the compactness ($M_p/R$) of the star. Direct measurements of the mass $M_p$ from pulsar timing therefore substantially improve constraints on the radius $R$ and the equation of state. Here we use observations spanning 26 years from \textit{Murriyang}, the 64-m Parkes radio telescope, to improve the timing model for this pulsar. Among the new precise measurements are the pulsar mass $M_p=1.418\pm 0.044$\,$M_{\odot}$, distance $D=156.96 \pm 0.11$\,pc, and orbital inclination angle $i=137.506 \pm 0.016^\circ$, which can be used to inform the X-ray pulse profile models inferred from NICER observations. We demonstrate that these results are consistent between multiple data sets from the Parkes Pulsar Timing Array (PPTA), each modeled with different noise assumptions. Using the longest available PPTA data set, we measure an apparent second derivative of the pulsar spin frequency and discuss how this can be explained either by kinematic effects due to the proper motion and radial velocity of the pulsar or excess low-frequency noise such as a gravitational-wave background.
\end{abstract}

\section{Introduction} \label{sec:intro}

The equation of state (EoS) of neutron stars is an important probe of the physics at supranuclear densities and low (nonrelativistic) temperatures that are inaccessible to laboratory experiments \citep{Ozel+16}. Precision radio-frequency timing of millisecond pulsars (MSPs) can place stringent constraints on the EoS through neutron star mass measurements \citep[e.g.][]{Demorest+10, Cromartie+20, Fonseca+21}.

Pulsar timing arrays (PTAs) regularly monitor a set of MSPs over years, primarily to act as a Galactic-scale detector of nanohertz-frequency gravitational waves \citep[e.g.][]{ng_gwb, epta_gwb, PPTA-DR3_gwb, cpta_gwb}. The long-duration PTA data sets are also ideally suited for studying the astrometry and orbits of MSPs, leading to tests of theories of gravity and the physics of neutron star interiors. The Parkes Pulsar Timing Array (PPTA) was the first such long-term PTA observing project \citep{Manchester+13}. One of the most important pulsars in the array is the nearest and brightest MSP, J0437$-$4715 \citep{Johnston+93}. The proximity, long timing baseline ($\gtrsim 20$\,yr) and high-precision arrival times for PSR~J0437$-$4715 provide one of the best opportunities to constrain the neutron star EoS, in combination with a radius measurement.

The Neutron Star Interior Composition Explorer \citep[NICER;][]{Gendreau+12, Gendreau+16} is a NASA telescope on the International Space Station that observes and times MSPs in soft X-rays ($\sim 0.3$\,-\,$3$\,keV). The spectral-timing pulse profile data from NICER is modeled to infer X-ray emission region geometry and relativistic effects that yield the neutron star mass and radius. NICER has already constrained the mass and radius for the isolated PSR~J0030+0451 \citep{Nicer_J0030_radius1, Nicer_J0030_radius2} and the more massive PSR~J0740+6620 \citep{Nicer_J0740_radius1, Nicer_J0740_radius2} by leveraging a mass constraint from radio-frequency pulsar timing \citep{Cromartie+20, Fonseca+21}. However, PSR~J0437$-$4715 remains a primary target for the mission because of its proximity, brightness, and the precise independent mass measurement enabled by the extremely accurate pulsar timing. The mass of PSR~J0437$-$4715 is near the peak of the neutron star mass distribution \citep{Ozel+16b}, meaning that a radius measurement will constrain the neutron star EoS at the crucial intermediate densities that describe much of the population. A significant challenge in determining the mass and radius of PSR~J0437$-$4715 from NICER data alone is the presence of a bright active galactic nucleus in the field that contributes additional background noise. The NICER models are improved further by prior knowledge of the pulsar distance and orbital inclination angle, which can also be inferred from PPTA observations \citep{Reardon+16}.

The times of arrival (ToAs) of MSPs reveal not just the deterministic physics captured by timing model parameters but also numerous stochastic processes \citep{NG_15yr_noise, PPTA-DR3_noise, EPTA_noise}. Examples of these processes include interstellar dispersion measure (DM) and scattering variations \citep[e.g.][]{Keith+13, Donner+20, Turner+21},  pulsar rotation irregularities \citep[e.g.][]{Shannon+10}, pulse jitter \citep[e.g.][]{Liu+12, Shannon+14, Lam+19, Parthasarathy+21}, radio-frequency interference (RFI), sudden and secular profile shape variations \citep[e.g.][]{Shannon+16, Jennings+24},  calibration errors in specific observing systems \citep[e.g.][]{vanStraten13, Lentati+16}, and nanohertz-frequency gravitational waves. PSR~J0437$-$4715, shows evidence for most of these noise processes. Determining precise and accurate measurements of the neutron star mass, distance, and inclination requires both accurate ToAs and noise models. 

Throughout the NICER mission, the modeling of PSR~J0437$-$4715 observations has been conducted using informative prior probability distributions derived from PPTA observations. In this Letter, we describe these PPTA data sets and noise models (Section \ref{sec:data}), the timing model parameters and inference technique (Section \ref{sec:models}), and our resulting measurements and derived quantities (Section \ref{sec:results}). Our results are discussed in Section \ref{sec:discussion} and we conclude in Section \ref{sec:conclusion}.

\section{Data sets} 
\label{sec:data}

\begin{figure*}
\centering
\includegraphics[width=\textwidth]{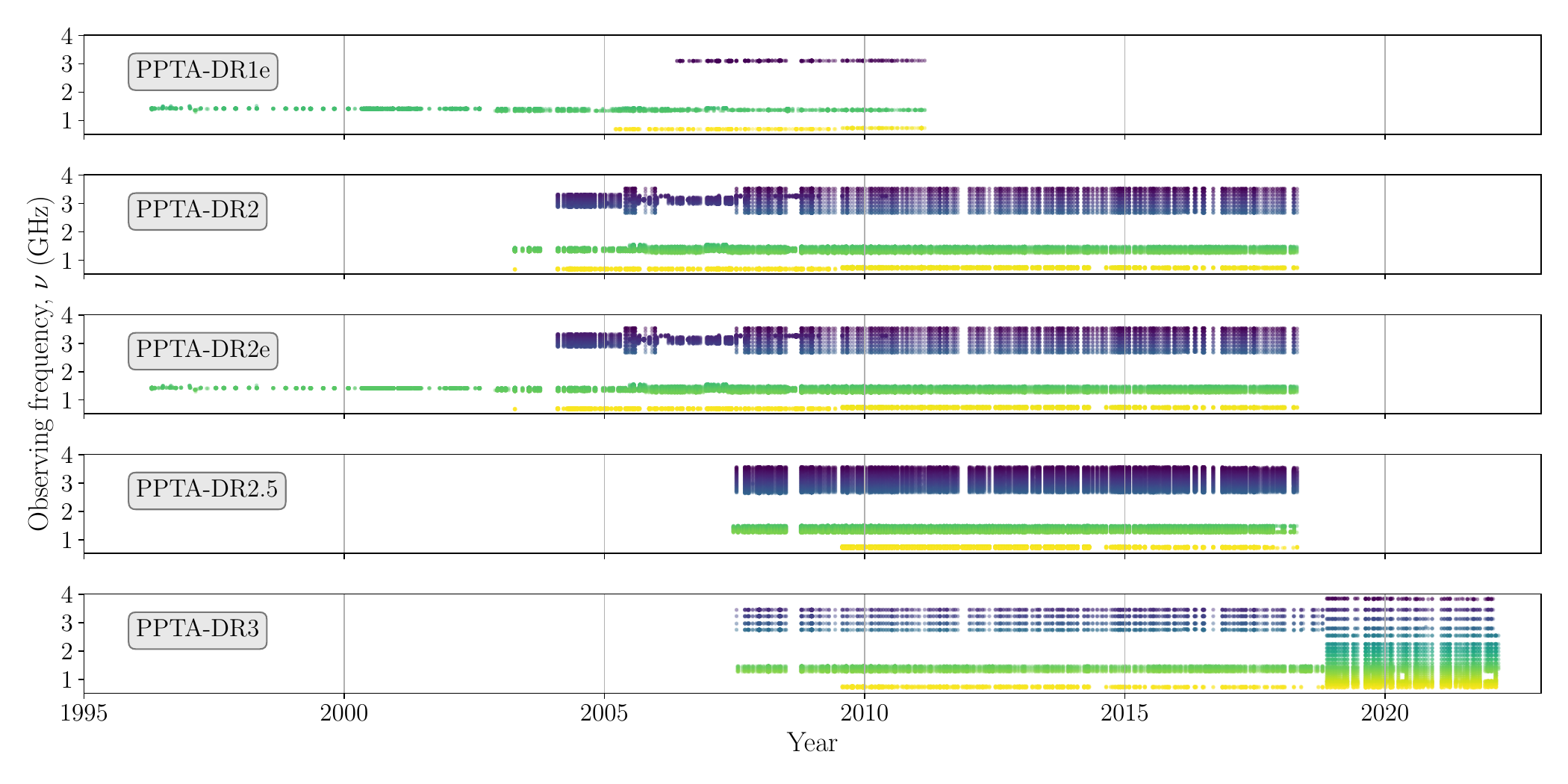}
\caption{Measured ToAs for PSR~J0437$-$4715 across multiple PPTA data sets. The colors of the points indicate the observing frequency on a linear scale. The data set is labelled on each panel, and described in Section \ref{sec:data}.}
\label{fig:observations}
\end{figure*}

Pulsar timing model parameters can change significantly between different analyses. For this work, we critically analyze different PPTA data sets to demonstrate that our results are robust. The measured ToAs as a function of time and frequency, for all data sets discussed in this section, are shown in Figure \ref{fig:observations}.

\subsection{PPTA first data release}
The PPTA commenced regular observations of a set of MSPs in 2004 and produced a first data release of 20 pulsars after six years of observations \citep{Manchester+13}. This first data release (PPTA-DR1) was combined with earlier ``legacy" observations of many pulsars taken under different projects, which for PSR~J0437$-$4715 dated back to 1996 \citep{Verbiest+08}. The combination of these data with PPTA-DR1 is referred to as the ``extended" first data release (PPTA-DR1e). A timing and noise analysis of the 20 MSPs in PPTA-DR1e, including PSR~J0437$-$4715, was conducted by \citet{Reardon+16} and included precise measurements of the pulsar mass, $M_p$, and distance, $D$. For PPTA-DR1 and all subsequent data releases to date, the pulsar was timed using the polarization invariant interval \citep{Britton00} to mitigate polarization calibration errors \citep{Manchester+13}. 

\subsection{PPTA second data release: Noise model and legacy data extension}
\label{sec:ppta-dr2}

The PPTA second data release (PPTA-DR2) is described in \citet{ppta_dr2} and included observations of 26 pulsars, spanning up to 14 years. The timing residuals were analyzed in detail for noise processes by \citet{ppta_dr2_noise}. Despite using the most comprehensive noise model ever created for a pulsar, the normalized and noise-subtracted timing residuals for PSR~J0437$-$4715 failed to pass any of the statistical tests used to assess if the model fully characterizes the noise. These challenges have been noted for PSR~J0437$-$4715 during numerous earlier analyses \citep[e.g.][]{Verbiest+08, vanStraten13, Lentati+14, Lentati+16, Reardon+16}.

For a detailed description of the PPTA-DR2 noise models, see \citet{ppta_dr2_noise}. In brief, the noise models include white-noise parameters (that account for uncorrelated instrumental noise and pulse jitter) and Gaussian random processes with a power-law power spectral density \citep[PSD; e.g.][]{vanHaasteren+13, Lentati+13}. These Gaussian processes are used to account for achromatic noise (for example spin noise and gravitational waves), DM variations (scaling in amplitude $\propto \nu^{-2}$, for observing frequency $\nu$), scattering variations (scaling with $\nu^{-4}$), and excess achromatic noise in specific observing bands (band noise) and observing systems (system noise). The parameters inferred for each such process are the PSD power law amplitude, $A$, and spectral exponent, $\gamma$.

The PPTA-DR2 data set was then extended with the legacy observations in \citet{Reardon+21}, forming PPTA-DR2e, for all pulsars except PSR~J0437$-$4715. The timing models for these pulsars were updated, with the parameters inferred in the presence of the \citet{ppta_dr2_noise} noise models with least-squares regression in the pulsar timing package \textsc{tempo2} \citep{Hobbs+06, Edwards+06}. The white-noise parameters, as well as $A$ and $\gamma$ for each Gaussian process, were fixed at their inferred maximum likelihood values. The Gaussian processes are implemented in the timing model as a Fourier series with $N$ component frequencies, from $f_{\rm min} = 1/T_{\rm span}$ to $f_{\rm max} = N/T_{\rm span}$, for total data set time span, $T_{\rm span}$. The amplitudes of the Fourier basis functions are included in the timing model but constrained to follow the PSD described by $A$ and $\gamma$. The noise models from PPTA-DR2 were extrapolated for use with PPTA-DR2e by increasing the number of Fourier components, $N$, such that the highest-frequency component, $f_{\rm max}$, was similar for both data sets \citep{Reardon+21}. 

For this work, we repeated this process for PSR~J0437$-$4715, to produce a 22-year PPTA-DR2e data set and an extrapolated noise model. The timing model in this data set assumes the DE436 solar system ephemeris model \citep[an updated version of DE430;][]{Folkner+14, Park+21} since this was assumed for the noise analysis \citep{ppta_dr2_noise}. As with PPTA-DR2, the noise model whitens the residuals on many timescales, but the noise-subtracted residuals remain non-Gaussian and heteroscedastic. This is likely due to changes in the noise statistics of the multiple observing systems over the two decades of PPTA observations. While the effect of such noise model deficiencies on the timing model parameters are expected to be small, based on attempts to model the timing residuals with a non-Gaussian likelihood \citep{Lentati+14}, we endeavored to improve the residuals before these parameters were adopted for the NICER mission.

\subsection{Reprocessing the PPTA second data release: PPTA-DR2.5}

To better understand the incomplete model of noise in PPTA-DR2, we produced an \textit{ad hoc} data set by reprocessing the PPTA-DR2 pulse profiles of PSR~J0437$-$4715, with the goal to first produce the highest-quality ToAs and uncertainties. We refer to this data set as ``PPTA-DR2.5" as it was produced specifically to achieve new timing model measurements for the NICER mission, prior to the publication of the PPTA third data release (PPTA-DR3).

The published PPTA-DR2 data set included calibrated (polarization and flux density) and RFI-excised pulse profiles with 32 frequency channels and eight subintegrations in time. We developed a pipeline extension to the PPTA-DR2 data reduction pipeline to reprocess these pulse profiles. This extension was used in PPTA-DR3, and it is therefore also discussed in \citet[][Section 2.2]{PPTA-DR3_data}. Below we summarize the key pipeline changes and two differences between our \textit{ad hoc} PPTA-DR2.5 (produced only for PSR~J0437$-$4715) and the published PPTA-DR3:

\begin{itemize}
    \item Both PPTA-DR2.5 and PPTA-DR3 only include observations from the more modern signal processing systems (backends) as the earlier systems contribute unidentified excess noise to the ToAs \citep{ppta_dr2_noise}. The data sets therefore start in July 2007 with observations recorded with the \textsc{PDFB2} backend. PPTA-DR3 additionally includes $\sim 3.3$\,years of observations from the ultrawide-band low (UWL) receiver \citep{Hobbs+20}. 
    \item PPTA-DR2.5 retained all 32 frequency channels of the pulse profiles, while PPTA-DR3 was reduced to four channels per observation (or per sub-band for UWL observations). These differences are apparent in Figure \ref{fig:observations}.
    \item For each combination of observing receiver and backend, we created a separate template profile by summing the brightest 100 observations in time and smoothing the resulting 32-channel profile with wavelet smoothing using the \textsc{psrsmooth} of the \textsc{psrchive} software package \citep{vanStraten+12}. For PPTA-DR3, a single analytical wide-band profile \citep[derived from UWL observations;][]{Curylo+23} was used.
    \item Each channel in the pulse profile was timed with the corresponding channel from the template, with the ToA and uncertainty measured with the Fourier-domain Monte Carlo method in \textsc{pat} \citep{vanStraten+12}.
\end{itemize}

The noise in  PPTA-DR2.5 was modeled using \textsc{temponest} \citep{temponest}. Power-law models were used to describe the DM variations, achromatic red noise, and band noise in two bands: $\nu < 1$\,GHz and $1\,{\rm GHz} < \nu < 2$\,GHz. White-noise parameters were inferred for each combination of receiver and backend. We found that PPTA-DR2.5 did not require any additional timing model parameters to account for profile evolution with frequency, while PPTA-DR3 did \citep{PPTA-DR3_data}.

Following this noise analysis of PPTA-DR2.5, the frequency-averaged, whitened, and normalized residuals were found to be statistically white and Gaussian distributed using the same tests as in \citet{ppta_dr2_noise, Reardon+21}. Inclusion of ToAs from earlier observing systems caused these tests to fail, which motivated the use of only these observing systems for PPTA-DR3.

\subsection{PPTA third data release: PPTA-DR3}

\citet{PPTA-DR3_data} describes the third PPTA data release, which was analyzed for noise processes in \citet{PPTA-DR3_noise}, using the \textsc{enterprise} package \citep{Enterprise}. We do not further modify the noise model or ToAs from those already published. The noise analysis for PPTA-DR3 assumed the DE440 JPL solar system ephemeris \citep{Park+21}. 

\begin{figure*}
\centering
\includegraphics[trim=0cm 0.2cm 0cm 0cm,clip=true,width=\textwidth]{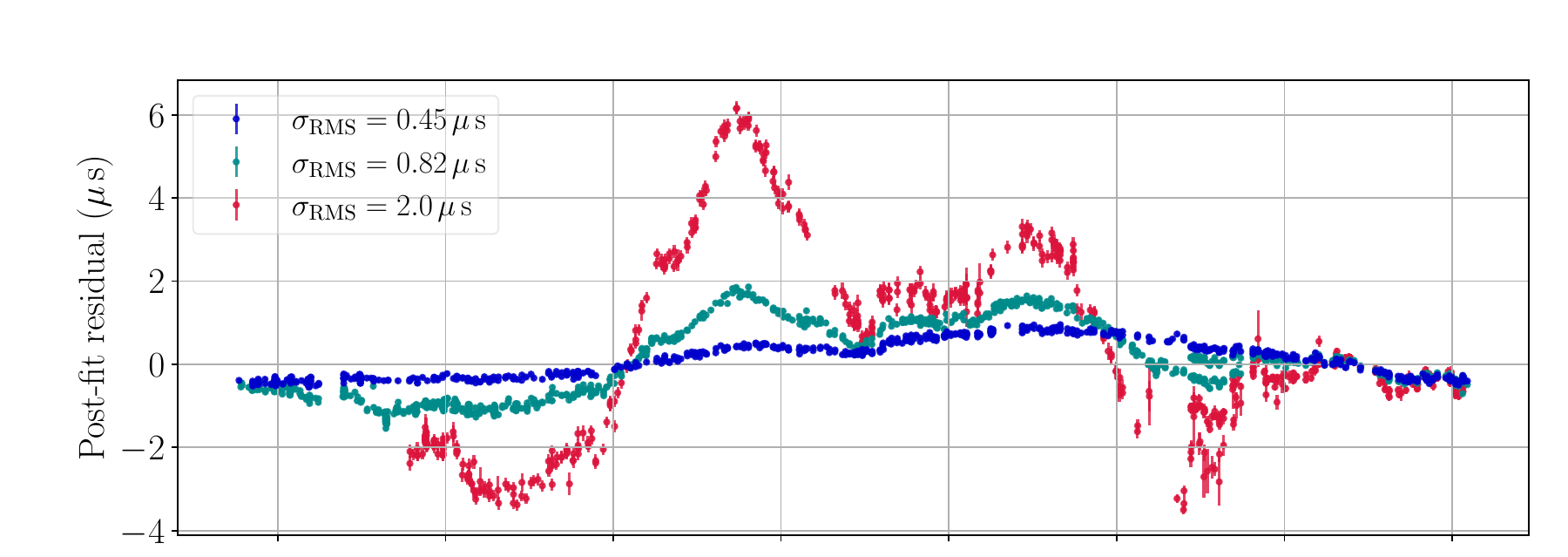}
\includegraphics[trim=0cm 0cm 0cm 0.15cm,clip=true,width=\textwidth]{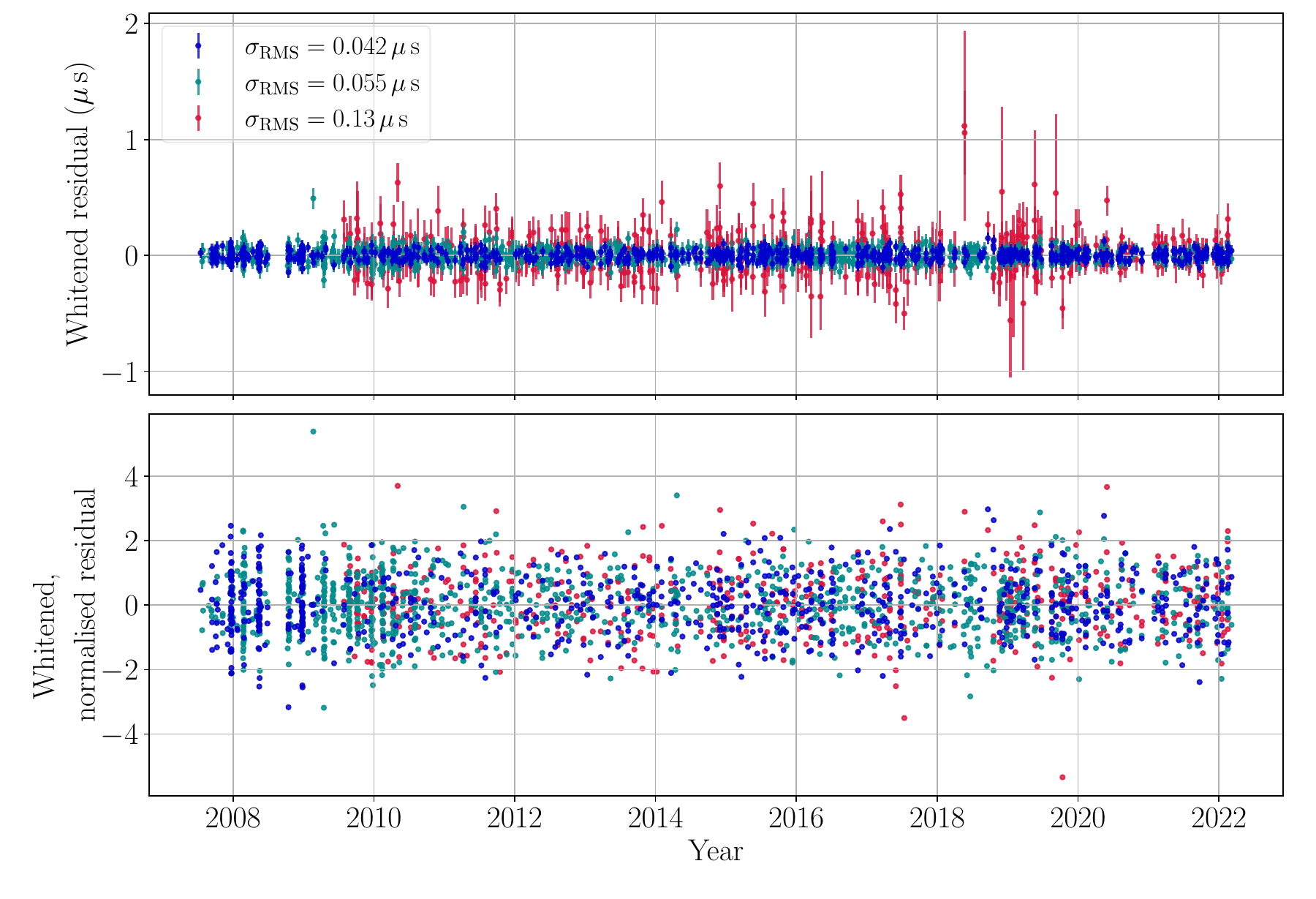}
\caption{Band-averaged timing residuals for PSR~J0437$-$4715 from PPTA-DR3. Residuals at observing frequencies near 800\,MHz, 1400\,MHz, and 3100\,MHz are shown in red, teal, and blue, respectively. Top panel shows post-fit residuals with all modeled Gaussian random processes present, including the dominant DM variations that increase in amplitude $\propto \nu^{-2}$. The middle panel shows the residuals after removal of the maximum likelihood realizations of these Gaussian processes. The bottom panel shows these noise-subtracted residuals divided by the uncertainty. The top and middle panels have labels that show the weighted rms of the residuals in each band.
}
\label{fig:residuals}
\end{figure*}

The timing residuals for PSR~J0437$-$4715 from PPTA-DR3 are shown in Figure \ref{fig:residuals}. The frequency-averaged, noise-subtracted (whitened), and normalized residuals in the bottom panel pass the same tests as PPTA-DR2.5 and so are also statistically white and Gaussian distributed as expected from a complete noise model. 

\section{Timing model}
\label{sec:models}

PSR~J0437$-$4715 is in a $5.74$-day, near-circular orbit (eccentricity, $e\sim 1.9\times 10^{-5}$) with a helium white dwarf companion \cite[][]{Bell+93}. The brightness and proximity of the pulsar, as well as the geometry of the orbit, are favorable for measuring useful relativistic and kinematic effects from precisely measured ToAs. The timing model implemented in \textsc{tempo2} is described in detail in \citet{Edwards+06}, with the post-Newtonian orbital mechanics derived in \citet{DD}. We use the T2 binary model from \textsc{tempo2}, which is an extension of the model described in \citet{Damour+92} to include kinematic effects (e.g. Section \ref{sec:aop}). Here we provide a summary of the most important timing model parameters relevant to this work.

\subsection{Shapiro Delay}

General relativity predicts that spacetime can be measurably compressed around massive and compact objects. As radio waves from PSR~J0437$-$4715 propagate through the compressed spacetime surrounding the white dwarf companion, they are delayed relative to propagation through uncompressed spacetime. This is the well-known Shapiro delay \citep{Shapiro64}, and it is observed most readily in binary MSPs with orbits that are observed near edge-on (highly inclined). It is parameterized in the timing model of PSR~J0437$-$4715 by the companion mass $M_c$ and shape $s = \sin{i}$, where $i$ is the orbital inclination angle. A precise measure of $M_c$ and $\sin{i}$ therefore yields the pulsar mass $M_p$ through the binary mass function. However, $i$ can also be measured independently through kinematic effects that change the projected shape of the orbit on the sky, thus eliminating the strong covariance between $M_c$ and $i$ due to Shapiro delay.

\subsection{Annual-orbital parallax}
\label{sec:aop}

For the nearest and most precisely timed binary pulsars, it is possible to measure annual variations in the projected shape of the orbit. This is the annual-orbital parallax and leads to variations in the projected semi-major axis, $x = a\sin{i}$ (where $a$ is the semi-major axis), and longitude of periastron $\omega$ \citep{Kopeikin95}. In addition, secular variations in $x$ and $\omega$ originate from the proper motion of the binary slowly altering the projection of the orbit \citep{Kopeikin96}. These effects are parametrized in \textsc{tempo2} by the orbital inclination angle $i$ (which simultaneously describes the Shapiro delay shape) and the longitude of ascending node, $\Omega$. Thus, these ``Kopeikin terms" account for the annual-orbital parallax, the kinematic contributions to the time derivatives of $x$ and $\omega$, and also tightly constrain the Shapiro delay. 

Additional contributions to the observed time derivatives of $x$ and $\omega$ can arise, for example, from relativistic effects, or spin-orbit coupling. For PSR~J0437$-$4715, we detect a significant relativistic periastron advance, $\dot{\omega}$, and include this in the binary model. However, the only measurable component to $\dot{x}$ is the kinematic term that is accounted for with the Kopeikin terms; thus, we do not include an additional $\dot{x}$ in the model. 

\subsection{Shklovskii and parallax distances}

The proper motion of a pulsar not only changes the projection of the orbit but is also associated with an apparent radial acceleration that causes an increasing Doppler shift to the spin and orbital periods. This Shklovskii effect \citep{Shklovskii70} can be used to derive the pulsar distance from a measurement of the resulting orbital period derivative $\dot{P}_b^{\rm shk}$, with \citep{Bell+96}
\begin{equation}
    D_{\rm shk} = \frac{c}{\mu^2}\frac{\dot{P}_b^{\rm shk}}{P_b},
\end{equation}
where $\mu$ is the magnitude of the proper motion, $P_b$ is the orbital period, and $c$ is the speed of light. However, the observed orbital period derivative, $\dot{P}_b^{\rm obs}$, has additional contributions, including  from the radiation of gravitational waves by the system, $\dot{P}_b^{\rm GW}$, and from differential acceleration in the Galaxy, $\dot{P}_b^{\rm Gal}$. The Shklovskii component can be computed by correcting for these effects
\begin{equation}
    \dot{P}_b^{\rm shk} = \dot{P}_b^{\rm obs} - \dot{P}_b^{\rm GW} - \dot{P}_b^{\rm Gal}.
\end{equation}

We use \textsc{galpy} \citep{Bovy15} to measure and correct for the Galactic acceleration term, $\dot{P}_b^{\rm Gal}$. We propagate uncertainties in this correction by assuming a 3\% uncertainty for the circular differential acceleration \citep{Bovy+12} and 10\% for the vertical acceleration in the Galactic potential \citep{Holmberg+04}, finding $\dot{P}_b^{\rm Gal} = -(2.24 \pm 0.20)\times 10^{-14}$. The expected $\dot{P}_b$ from gravitational-wave emission is two orders of magnitude smaller, $\dot{P}_b^{\rm GW} \sim -3.2\times 10^{-16}$.

An independent distance can be found from the inverse of parallax, $\pi$, which in pulsar timing is a detection of the time delays of curved wave fronts (with respect to plane waves) arriving from a finite distance. However, the sensitivity of a pulsar timing data set to parallax  $D_{\pi} 
 \propto T_{\rm span}^{1/2}$ does not grow as rapidly with time as $D_{\rm shk}$, which improves $ \propto T_{\rm span}^{5/2}$.  

\subsection{Radial velocity}
\label{sec:radV_intro}

The radial motion of the pulsar causes an apparent second derivative of the pulsar spin period, $\ddot{f}$ \citep{vanStraten03, Liu+18}. It also causes a proper motion derivative \citep{vanStraten03}, which will have a smaller effect on the timing residuals. However, the observed $\ddot{f}$ includes contributions from multiple other sources,
\begin{equation}
\label{eqn:f2}
    \ddot{f}_{\rm obs} = \left(1 - \frac{v_r}{c}\right)\ddot{f}_{\rm int} + \ddot{f}_{\rm shk} + \ddot{f}_{\parallel} + \ddot{f}_{\perp} + \ddot{f}_{\rm acc} + \ddot{f}_{\rm jerk} + \ddot{f}_{\rm red},
\end{equation}
where $v_r$ is the radial velocity; $\ddot{f}_{\rm int}$ comes from the intrinsic spin-down of the pulsar; $\ddot{f}_{\rm shk}$ originates from the Shklovskii effect acting on the intrinsic $\dot{f}$; $\ddot{f}_{\parallel}$ and $\ddot{f}_{\perp}$ are additional Shklovskii terms that depend on the radial velocity and transverse acceleration, respectively; and $\ddot{f}_{\rm acc}$ and $\ddot{f}_{\rm jerk}$ depend, respectively, on differential acceleration and jerk \citep[for details on these terms, see][]{Liu+18}. $\ddot{f}_{\rm red}$ is an additional noise component due to unmodeled low-frequency (red) noise in the timing residuals.

For PSR~J0437$-$4715, the dominant term (assuming an accurate noise model, $\ddot{f}_{\rm red} = 0$) in Equation \ref{eqn:f2} is $\ddot{f}_{\parallel}$, which is proportional to the radial velocity. Under typical spin-down scenarios, the expected intrinsic $\ddot{f}_{\rm int}$ for low-magnetic-field MSPs is small. This and the other terms in Equation \ref{eqn:f2} are estimated to contribute $\lesssim 1\%$ to $\ddot{f}_{\rm obs}$ \citep[][Table 1]{Liu+18}. For this estimate, the jerk and acceleration terms were calculated for the Galactic potential only. Interactions with field stars or a second companion in a wide orbit could potentially increase these components, but this is unlikely to be significant for PSR~J0437$-$4715. One Galactic (field) MSP was found to be in a wide binary orbit that manifested as a significant $\ddot{f}$ due to gravitational jerk \citep{Bassa+16, Kaplan+16}. However, unlike PSR~J0437$-$4715, this pulsar also had an apparent $\dot{f}$ that was inconsistent with spin-down alone because of a significant contribution from gravitational acceleration, and the wide orbit companion was identified optically. 

The radial velocity is then calculated assuming $\ddot{f}_{\parallel} \approx \ddot{f}_{\rm obs}$ with
\begin{equation}
\label{eqn:radial_v}
    v_r = \frac{\ddot{f}_{\parallel}}{f_0} \frac{c}{3 \mu^2}.
\end{equation}
However, covariance between $\ddot{f}$ and red noise processes makes the measurement challenging, particularly if the noise has a steep spectrum \citep{Liu+19, Keith+23}. Indeed leakage of red noise into the $\ddot{f}$ measurement due to model misspecification, $\ddot{f}_{\rm red}$, may be one of the dominant terms in Equation \ref{eqn:f2}. The precision in $\ddot{f}$, and hence $v_r$, is estimated to improve $\propto T_{\rm span}^{\sim 2}$ for PSR~J0437$-$4715, after accounting for covariance with the achromatic red noise \citep{Liu+19}, which has spectral exponent $\gamma\sim -3$ \citep{ppta_dr2_noise, PPTA-DR3_noise}. If the red noise in PSR~J0437$-$4715 becomes dominated by the steeper ``common-spectrum" process observed using PPTA-DR3 \citep[which may be a gravitational-wave background;][]{PPTA-DR3_gwb}, then the precision in $\ddot{f}$ would begin to improve more slowly, $\propto T_{\rm span}^{\sim 1.5}$, for $\gamma\sim -3.9$. With the present sensitivity of PPTA data sets, the magnitude of $v_r$ can only be estimated under an assumed noise model. Confirmation of $v_r$ from pulsar timing will be possible upon measurement of the proper motion derivative (requiring $\gtrsim 30$\,years of timing baseline), which has a purely geometrical origin \citep{vanStraten03}.

\subsection{Parameter inference}

Bayesian inference is commonly used for noise modeling of MSPs but can be computationally expensive for large data sets like those of the PSR~J0437$-$4715 ToAs. This is especially true when greatly increasing the dimensionality of the parameter space by choosing to sample over timing model parameters. Fortunately, for the case of timing model parameters that are measured with high significance, generalized least-squares fitting of the linearized timing model in codes such as \textsc{tempo2} is an efficient alternative and provides accurate measurements and uncertainties, provided the noise model is well specified and the fixed noise parameters (e.g., the amplitude $A$ and spectral exponent $\gamma$ of a power-law process) are uncorrelated with the timing model parameters. We verified this using \textsc{temponest} to sample $M_c$, $\dot{P}_b$, and $i$ simultaneously with the noise model for PPTA-DR2.5, finding negligible correlations between the parameters. We further verified that the results of the least-squares fit in \textsc{tempo2} (including the noise model) were consistent with the Bayesian posteriors \citep[see also][]{temponest}. 

For the analysis of all data sets, we therefore used this generalized least-squares method after validation of the noise models. The Gaussian process components of the noise models are realized simultaneously with the timing model fit, as described in Section \ref{sec:ppta-dr2} for the PPTA-DR2e analysis of other MSPs in the PPTA \citep{Reardon+21}. To derive further quantities from our measured timing model parameters (e.g. $M_p$ and $D$), we perform a Monte Carlo simulation of our measurements by drawing 1 million samples from Gaussian distributions described by the timing model parameter values and their uncertainties. For each set of parameter samples drawn, we derived the relevant quantity. The quoted derived quantities in Section \ref{sec:results} represent the mean and standard deviation of the distribution of these derived quantities.

\section{Results} \label{sec:results}

We present timing model parameter measurements derived from the two most sensitive data sets: PPTA-DR2e and PPTA-DR3. As the data set with the longest timing baseline, PPTA-DR2e is more sensitive to parameters that grow rapidly with $T_{\rm span}$. PPTA-DR3 has higher-quality ToAs from modern backends and the UWL receiver, as well as a complete noise model suitable for accurate measurements of effects at the smaller timescales, like the Shapiro delay and annual-orbital parallax. The parameters inferred from these data sets are presented in Table \ref{tab:params}.

The DM is taken from PPTA-DR2.5, which did not require the use of parameters that describe frequency-dependent (FD) pulse profile evolution \citep{NG11}. DM derivative terms, magnetospheric exponential dip parameters \citep{ppta_dr2_noise}, observing system jumps, and FD parameters are not presented here because we consider these part of the noise models. The ToAs from each data set and the complete pulsar ephemerides, which includes system jumps, noise model parameters, and other ancillary \textsc{tempo2} parameters, are available from the CSIRO Data Access Portal \citep{DAP} at \url{https://doi.org/10.25919/20rx-5f63}.  

\begin{table*}
\centering
\caption{Measured timing model parameters and derived quantities for PSR~J0437$-$4715 from PPTA-DR2e and PPTA-DR3. Values in parentheses are the 1-$\sigma$ standard errors on the last quoted decimal place. Values in bold are the most precise of the two data sets. $^{\dagger}$The measured DM  comes from PPTA-DR2.5, which did not require profile frequency evolution parameters that are highly covariant with DM. $^{\ddagger}$The radial velocity is derived assuming no excess noise contributes to the measured $\ddot{f}$, as discussed in Section \ref{sec:radV}.}
\begin{tabular}{lcc}
\hline\hline
\multicolumn{3}{c}{Observation Properties} \\
\hline\hline
& PPTA-DR2e & PPTA-DR3 \\ 
\hline
Reference epoch (MJD) \dotfill & 55486 & 55486 \\
Reference terrestrial time standard \dotfill & BIPM(2018) & BIPM(2020) \\
Solar system ephemeris \dotfill & DE436 & DE440 \\
Observing span (yrs) \dotfill & 22.016 & 14.641  \\
Number of observations \dotfill & 6396 & 3161 \\
\textsc{tempo2} Binary model \dotfill & T2 & T2 \\
\hline\hline
\multicolumn{3}{c}{Measurements} \\
\hline\hline
& PPTA-DR2e & PPTA-DR3 \\ 
\hline
Spin frequency, $f$ (s$^{-1}$)\dotfill & $\mathbf{173.6879456649435(2)}$ & 173.6879456649439(4) \\ 
Spin frequency derivative, $\dot{f}$ ($10^{-15}\,$s$^{-2}$)\dotfill & $\mathbf{-1.7283626(13)}$ & $-$1.728367(3) \\ 
Spin frequency second derivative, $\ddot{f}$ ($10^{-30}\,$s$^{-3}$)\dotfill & $\mathbf{-61(15)}$ & -- \\ 
Right ascension, $\alpha$ (hh:mm:ss, J2000) \dotfill & $\mathbf{04}$\bf{:}$\mathbf{37}$\bf{:}$\mathbf{15.9284039(2)}$ & 04:37:15.9284042(4)  \\ 
Declination, $\delta$ (dd:mm:ss, J2000) \dotfill & $\mathbf{-47}$\bf{:}$\mathbf{15}$\bf{:}$\mathbf{09.303707(2)}$ & $-$47:15:09.303700(4) \\
Dispersion measure, DM (pc\,cm$^{-3}$) \dotfill & 2.64539(3)$^{\dagger}$ & 2.64539(3)$^{\dagger}$ \\
Proper motion in $\alpha$, $\mu_\alpha \cos\delta$ (mas\,yr$^{-1}$)\dotfill & $\mathbf{121.4429(5)}$ & 121.4420(6) \\ 
Proper motion in $\delta$, $\mu_\delta$ (mas\,yr$^{-1}$)\dotfill & $\mathbf{-71.4715(5)}$ &  $-$71.4717(7) \\ 
Parallax, $\pi$ (mas)\dotfill & $\mathbf{6.43(4)}$ & 6.43(5) \\ 
Orbital period (days) \dotfill & $\mathbf{5.74104582(16)}$ & 5.74104635(19) \\ 
Epoch of periastron (MJD) \dotfill & $\mathbf{54530.17223(19)}$ &  54530.1722(2) \\ 
Projected semi-major axis (lt-s) \dotfill &  $3.36671463(3)$ & $\mathbf{3.36671466(3)}$ \\ 
Longitude of periastron (deg) \dotfill & $\mathbf{1.358(12)}$ & 1.359(13) \\ 
Eccentricity ($\times 10^{-5}$) \dotfill & $1.91819(11)$  & $\mathbf{1.91805(10)}$ \\  
Orbital period derivative, $\dot{P_b}$ ($\times 10^{-12}$)\dotfill & $\mathbf{3.7329(15)}$ & 3.733(3)  \\  
Periastron advance, $\dot{\omega}$ (deg\,yr$^{-1}$)\dotfill & $\mathbf{0.0134(6)}$ & 0.0155(8)  \\ 
Companion mass, $M_c$ ($M_\odot$)\dotfill & $0.224(4)$  & $\mathbf{0.221(4)}$  \\  
Longitude of ascending node, $\Omega$ (deg)\dotfill & $208.5(9)$  & $\mathbf{208.3(8)}$  \\  
Orbital inclination, $i$ (deg)\dotfill & $137.504(18)$  & $\mathbf{137.506(16)}$ \\  
\hline\hline
\multicolumn{3}{c}{Derivations} \\
\hline\hline
& PPTA-DR2e & PPTA-DR3 \\ 
\hline
Pulsar mass (Shapiro delay), $M_p$ ($M_\odot$) \dotfill & $1.446(45)$  &  $\mathbf{1.418(44)}$  \\  
Pulsar distance (Shklovskii effect), $D_{\rm shk}$ (pc) \dotfill &  $\mathbf{156.96(11)}$ & $156.98(15)$ \\  
Pulsar distance (parallax), $D_{\pi}$ (pc) \dotfill &  $\mathbf{155.4(9)}$ & $155.4(12)$  \\  
Radial velocity, $v_r$ (km\,s$^{-1}$) \dotfill &  $\mathbf{-75(18)}$$^{\ddagger}$ & -- \\  
\hline
\end{tabular}
\label{tab:params}
\end{table*} 

\begin{figure*}
\centering
\includegraphics[width=\textwidth]{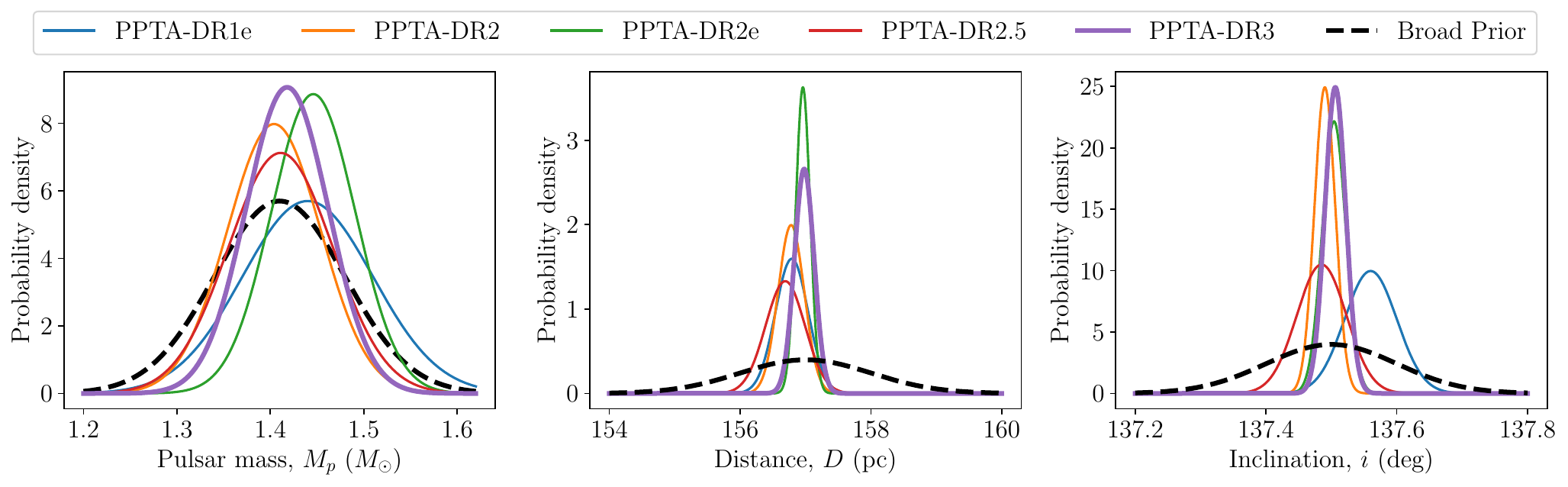}
\caption{Gaussian probability densities for the pulsar mass, distance, and orbital inclination measurements from various PPTA data sets, relative to the broad prior adopted for NICER early analyses and model selection.}
\label{fig:nicer}
\end{figure*}

\subsection{Precision measurements of neutron star mass, distance, and orbital inclination} \label{sec:measurements}

The timing model parameters of most importance for this work ($M_c$, $i$, and $\dot{P}_b$) are measured with high significance, are uncorrelated (between other timing model parameters and the noise processes), and are Gaussian distributed. The NICER models use our PPTA-DR3 measurements of $M_p=1.418\pm0.044$\,$M_{\odot}$, $D=156.98\pm0.15$\,pc, and $i=137.506\pm0.016$\,degrees, derived from these key parameters (with a Monte Carlo simulation), as Gaussian priors.

For the purpose of model selection and early testing in mid-2019, the inference of NICER observations adopted Gaussian priors (``Broad prior" in Figure \ref{fig:nicer}) that were centered near the best available measurement at the time (from PPTA-DR2.5). The standard deviation of the priors were broader for $i$ and $D$ as these terms were thought to be most susceptible to errors in noise models. These values are listed with measurements from all other data sets in Table \ref{tab:priors} and visualized in Figure \ref{fig:nicer}.

The orbital inclination is assumed to be equal to the inclination of the spin axis of the neutron star itself. MSPs are formed by mass transfer from a companion star that works to align the spin angular momentum with the orbital angular momentum over $\sim 100\,$Myr \citep{Tauris+99}. One NICER analysis of PSR~J0740+6620 assumed a $\pm5^\circ$ uniform prior around the timing model measurement of $i$ to allow for slight deviations from this expected alignment \citep{Nicer_J0740_radius2}. It was determined, however, that the data was insensitive to the precise choice of $i$ prior. 

\section{Discussion}
\label{sec:discussion}

Here we discuss the consistency of the parameter measurements from various data sets, including comparisons with the broader literature, and our derivation of the radial velocity. 

\subsection{Distance measurements}

Knowing MSP distances to high precision is valuable to PTA searches for continuous gravitational waves. If the distance error is much less than the wavelength of the gravitational wave, the pulsar term can be predicted, forming a coherent baseline for the array. We have inferred a high-precision distance from the Shklovskii effect on the orbital period derivative, $D_{\rm shk} = 156.96 \pm 0.11$\,pc, after correcting for differential Galactic acceleration. There is a $1.7\sigma$ discrepancy with the more uncertain distance inferred from the timing parallax, $D_{\pi} = 155.4\pm 0.9$\,pc. The independent astrometric distance from very-long-baseline interferometry \citep[VLBI;][]{Deller+08}, $D_{\rm VLBI} = 156.3 \pm 1.3$\,pc does not resolve the tension. A new distance constraint from an analysis of scintillation arcs, $D_{\rm scint}=158.1\pm1.5$\,pc, also differs from the timing parallax distance by $\sim 1.5\sigma$ and is more consistent with the $\dot{P}_b$ distance (D. Reardon et al. 2024, \textit{submitted}).

\subsection{Second spin frequency derivative}
\label{sec:radV}

Using PPTA-DR2e and the noise model of \citet{ppta_dr2_noise}, we measure a second derivative of the spin frequency, $\ddot{f} =  (-61 \pm 15) \times 10^{-30}\,$s$^{-3}$. This parameter could be interpreted as arising from kinematic effects (see Section \ref{sec:radV_intro}), or from excess low-frequency noise that was not characterized by the PPTA-DR2 noise models.

For the kinematic interpretation, we assume that the PPTA-DR2 noise model remains accurate when extrapolated to apply to the extended legacy data (therefore, $\ddot{f}_{\rm red} = 0$). We derive the radial velocity of the pulsar, $v_r = -75 \pm 18\,$km\,s$^{-1}$ from Equation \ref{eqn:radial_v}. The uncertainty accounts for covariance between $\ddot{f}$ and the Gaussian processes in the noise model, assuming fixed amplitudes and exponents. Although the noise model is incomplete for PPTA-DR2e \citep{ppta_dr2_noise}, it whitens the residuals at the longest timescales that are most relevant for obtaining an $\ddot{f}$ measurement. The noise models of \citet{ppta_dr2_noise} were determined simultaneously with the $\ddot{f}$ in the timing model so that the expected signature, if present, was not absorbed by the achromatic power-law Gaussian process.

To demonstrate the alternate $\ddot{f}$ interpretation of excess low-frequency noise, we instead extrapolate the common-spectrum stochastic process inferred from the 30 MSPs in PPTA-DR3 \citep{PPTA-DR3_gwb}. Assuming that this steeper-spectrum noise process (with $\gamma = -3.87$, $\log_{10} A = -14.5$) applies to PPTA-DR2e, we measure $\ddot{f} = (-19 \pm 45) \times 10^{-30}$\,s$^{-3}$, and therefore, a kinematic interpretation is not required. The constraint on radial velocity assuming this alternate noise model is $v_r = -24 \pm 40\,$km\,s$^{-1}$. Under this interpretation, the measured $\ddot{f}$ from PPTA-DR2e arises because the \citet{ppta_dr2_noise} noise model underestimates the achromatic noise PSD at the lowest frequencies.

A recent analysis of scintillation arcs using data from the MeerKAT radio telescope gives an independent estimate of the pulsar radial velocity by solving the three-dimensional geometry of the pulsar bow shock, $v_r = -44.5\pm4.3\,$km\,s$^{-1}$ (D. Reardon et al. 2024, \textit{submitted}), a difference of $1.6\sigma$ from the value derived from the kinematic $\ddot{f}$ interpretation. If this radial velocity is accurate, then $\ddot{f} = -36.3 \pm 3.5 \times 10^{-30}\,$s$^{-3}$ could be included in the timing model of PSR~J0437$-$4715 to improve the accuracy of recovered steep spectrum stochastic processes, including the putative gravitational-wave background. 

While the radial velocity and excess red noise are the most likely explanations for the measured $\ddot{f}$, it is also possible that some component is the consequence of reflex motion from a third, widely separated body in the system. To constrain the mass and orbital separation of a third body, we use the model presented in \citet{rasio94}, developed for the triple-system globular cluster MSP PSR~B1620$-$26 \citep{backer+93}.  We chose this model because it also provides the expected secular perturbations to the pulsar-white dwarf orbit. In this case, the third body mass $M_3$ and semimajor axis $a_3$ can be estimated as \citep{rasio94}
\begin{equation}
\ddot{f} = -\frac{f}{c} \frac{G^{3/2} M_3}{a_3^{7/2}} \sqrt{M_p + M_c + M_3} \cos(\omega + \phi) \sin i,   
\end{equation}
where $\phi$ is the position phase of the outer orbit relative to $\omega$ and the model assumes that the inner and outer orbits are coplanar. In the case that $M_3 \ll M_p + M_c$, we can approximate the relationship between $M_3$ and $a_3$ to be
\begin{eqnarray}
    M_3 && \approx  0.3  M_\earth \left(\frac{\ddot f}{61 \times 10^{-30}\,{\rm s}^{-3} }\right) \left( \frac{f}{173.7\,{\rm Hz}} \right)^{-1} \nonumber \\
     &&~~~\times \left(\frac{M_{\rm tot}}{1.67 M_\odot}\right)^{-1/2}  \left(\frac{a_3}{100\,{\rm AU}} \right)^{7/2},
\end{eqnarray}
where $M_{\rm tot}$ is the total mass of the system and we have assumed $\cos(\omega + \phi)=-1$. For a circular orbit with radius $100$\,AU, the orbital period would be $\approx 430$\,yr.

There is additional observational evidence that an outer companion would need to be low mass. Optical photometric and spectroscopic analysis of the companion(s) to PSR~J0437$-$4715 is consistent with a single helium white dwarf star lacking infrared excess that would be indicative of a massive nondegenerate companion \citep{Durant+12}.  However, an $M_3 \sim 0.03$\,$M_\odot$ L or T dwarf companion in a $2000$\,au, $40000$\,yr orbit would be less luminous in the infrared than the system white dwarf \citep{Charnay+18}. Such a companion would induce perturbations in $\dot{\omega}$ and $\dot{x}$ of $\mathcal{O}(10^{-11})$\,deg\,yr$^{-1}$ and $\mathcal{O}(10^{-20})$ respectively \citep{rasio94}. For any plausible $M_3$, these secular perturbations are orders of magnitude smaller than the current precision of these measurements.

\subsection{Consistency between PPTA data sets}

 We find that most model parameters in PPTA-DR2e and PPTA-DR3 are consistent within (1$\sigma$) uncertainties, with the exception of spin frequency $f$ (1.3$\sigma$), declination $\delta$ (1.5$\sigma$), proper motion in right ascension $\mu_\alpha$ (1.1$\sigma$),  and the highly covariant parameters $P_b$ and $\dot\omega$ (2.1$\sigma$). The discrepancy of $P_b$ and $\dot\omega$ likely originates from different noise model choices but has no significant effect on the other parameters of the model. The $\dot\omega$ value from PPTA-DR2e is more precise and more consistent with the expectation from general relativity than PPTA-DR3 (assuming the component masses from PPTA-DR3 and no contamination from kinematic effects as secular variations are accounted for with the Kopeikin terms). However, for these analyses, the $\dot\omega$ was not used to constrain the pulsar mass because the uncertainty is too large.

The Shapiro delay and annual-orbital parallax parameters are more precise with PPTA-DR3 because of the improved data quality. These effects gives the orbital inclination and the longitude of ascending node $\Omega$, which have also been measured using scintillation arcs in the PPTA-DR2 data set \citep{Reardon+20}. The value of $\Omega$ derived from these scintillation arcs differs by $2.2\sigma$, which is most likely to be explained by weakness in the scintillation arc curvature model, which assumes a stable interstellar scattering screen over many years.


\begin{table*}
\centering
\caption{Inferred pulsar mass, distance, and orbital inclination for PSR~J0437$-$4715, using various PPTA data sets.}
\begin{tabular}{lccc}
\hline\hline
 Data set & Pulsar mass, $M_p$ ($M_\odot$) & Distance, $D$ (pc) &  Orbital inclination, $i$ (deg) \\
\hline
PPTA-DR1e$^\dagger$ \dotfill & $1.44 \pm 0.07$ & $156.79 \pm 0.25$ & $137.56 \pm 0.04$ \\
PPTA-DR2 \dotfill & $1.404 \pm 0.050$ & $156.78 \pm 0.20$ & $137.490 \pm 0.016$ \\
PPTA-DR2e \dotfill & $1.446 \pm 0.045$ & $156.96 \pm 0.11$ & $137.504 \pm 0.018$ \\ 
PPTA-DR2.5 \dotfill & $1.411 \pm 0.056$ & $156.69 \pm 0.30$ & $137.485 \pm 0.038$ \\
PPTA-DR3 \dotfill & $1.418 \pm 0.044$ & $156.98 \pm 0.15$ & $137.506 \pm 0.016$ \\ 
\hline
Broad NICER prior \dotfill & $1.41 \pm 0.07$ & $157.0 \pm 1.0$ & $137.50 \pm 0.10$ \\
\hline
$^\dagger$ from \citet{Reardon+16}
\end{tabular}
\label{tab:priors}
\end{table*}

\section{Conclusion}
\label{sec:conclusion}

The precise ($\sim 3\%$ uncertainty) pulsar mass for PSR~J0437$-$4715 paves the way for a stringent constraint on the neutron star EoS. We find that the pulsar has a mass, $M_p = 1.418 \pm 0.044$\,$M_{\odot}$, near the center of the population of neutron star masses \citep{Ozel+16b}. A precise radius measurement from NICER, enabled by the radio-frequency timing mass, imposes a tight restriction on the allowed mass-radius relationship predicted by the EoS, particularly for this central mass range \citep{Choudhury+24, Rutherford+24}.

In our attempt to produce the most accurate pulsar mass measurement possible, we developed the first data sets with a complete noise model for PSR~J0437$-$4715 (PPTA-DR2.5 and PPTA-DR3). We compared the timing model parameters from multiple generations of PPTA data sets (with different noise models), and found that the parameters important for characterizing the physical properties of the system do not significantly vary. This is encouraging because data sets containing observations prior to 2007 are known to contain excess unmodeled noise.

Using the longest data set, PPTA-DR2e, we measure a significant second derivative of the pulsar spin frequency, $\ddot{f}$. We discussed two likely interpretations of this measurement, in the form of kinematic effects due to a pulsar radial velocity of $v_r = 75\pm18$\,km\,s$^{-1}$, or excess low-frequency noise such as from a gravitational-wave background. Additionally, we derived a constraint on the allowed mass and orbital separation of any third body in the case that some component of $\ddot{f}$ is due to reflex motion. Future analyses may consider using a modified Fourier basis \citep{Keith+23} to ensure robust radial velocity measurements, particularly if a gravitational-wave background begins to dominate the achromatic noise in longer data sets.

Future PPTA data sets could also be improved by employing new techniques such as matrix equation template matching for instrumental calibration, matrix template matching for ToA measurements \citep{vanStraten13, Rogers+24}, or wide-band timing \citep{Curylo+23}.

\section*{Acknowledgements}

Murriyang, the Parkes radio telescope, is part of the Australia Telescope National Facility (https://ror.org/05qajvd42) which is funded by the Australian Government for operation as a National Facility managed by CSIRO. We acknowledge the Wiradjuri people as the Traditional Owners of the Observatory site. Parts of this research were conducted by the Australian Research Council Centre of Excellence for Gravitational Wave Discovery (OzGrav), through project numbers CE170100004 and CE230100016.  R.M.S. acknowledges support through Australian Research Council Future Fellowship FT190100155. Work at NRL is supported by NASA. L.Z. is supported by ACAMAR Postdoctoral Fellowship and the National Natural Science Foundation of China (grant No. 12103069). 

\bibliography{references}

\end{document}